\documentclass[aps,prl,twocolumn,groupedaddress]{revtex4} 
 
\usepackage[utf8]{inputenc}  
\usepackage{graphicx,color} 
\usepackage{mathrsfs}
\usepackage{bm}
\usepackage{amsmath,amssymb,amsfonts}
\usepackage{amssymb}
\usepackage{ulem}
\usepackage{dsfont}
\usepackage{adjustbox}

\newcommand{\sumint}{\int\hspace{-14pt}\sum}

\def\bea{\begin{eqnarray}}
\def\eea{\end{eqnarray}}
\def\be{\begin{equation}}

\def\ee{\end{equation}}

\def\Put(#1,#2)#3{\leavevmode\makebox(0,0){\put(#1,#2){#3}}}

\begin{document}

\title{Microscopic positive-energy potential based on Gogny interaction}
\author{G. Blanchon$^{(1)}$, M. Dupuis$^{(1)}$, H. F.  Arellano$^{(1,2)}$, and N. Vinh Mau}
\affiliation{$^{(1)}$ CEA,DAM,DIF F-91297 Arpajon, France}
\affiliation{$^{(2)}$ Department of Physics - FCFM, University of Chile, Av. Blanco Encalada 2008, Santiago, Chile}

\begin{abstract}
{We present nucleon elastic scattering calculation based on Green's function formalism in the Random-Phase Approximation. 
For the first time, the Gogny effective interaction is used consistently throughout the whole calculation to account for the complex, 
non-local and energy-dependent optical potential. Effects of intermediate single-particle resonances are included and found to play a 
crucial role in the account for measured reaction cross section. Double counting of the particle-hole second-order contribution is 
carefully addressed. The resulting integro-differential Schr\"odinger equation for the scattering process is solved without localization 
procedures. The method is applied to neutron and proton elastic scattering from $^{40}$Ca. A successful account for differential and 
integral cross sections, including analyzing powers, is obtained for incident energies up to 30~MeV. Discrepancies at higher energies 
are related to much too high volume integral of the real potential for large partial waves. Moreover, this works opens the way for future 
effective interactions suitable simultaneously for both nuclear structure and reaction.}
\end{abstract}

\maketitle

Nuclear structure and nuclear reactions are two aspects of the same many-body problem, although in practice they are often addressed as 
different phenomena. A consistent, quantitative and predictive account for both is still a challenging open problem in nuclear physics. 
The description of nucleon-nucleus elastic scattering based solely on the nucleon-nucleon (NN) interaction is an important step forward 
toward this unification.

Depending on projectile energy and target mass, various strategies have been adopted in order to treat microscopically elastic 
scattering. Nuclear matter models \cite{hufner_72} provide reasonable descriptions of nucleon elastic scattering at incident 
energies above 50~MeV \cite{dupuis_06}, even up to $\sim$1~GeV \cite{arellano_02}. The Resonating Group Method within the No-Core Shell 
Model, has successfully described nucleon and deuteron scattering from light nuclei \cite{quaglioni_08}. These models 
have recently been extended to include three-nucleon forces for nucleon scattering from $^{4}$He \cite{hupin_13}. The Green's Function 
Monte Carlo method has been used to describe elastic scattering from $^{4}$He \cite{nollett_07}. These models yield encouraging 
results but are still restricted to light targets at low energies. The Self-Consistent Green's Function (SCGF) method has been 
applied to microscopic calculation of the optical potentials for proton scattering from $^{16}$O \cite{dussan_11,barbieri_05}. The 
coupled-cluster theory has been applied to proton elastic scattering from $^{40}$Ca \cite{hagen_12}. These last two methods are 
limited to closed-shell nuclei. Work on Gorkov-Green's function theory is in progress to extend SCGF to nuclei around closed-shell 
nuclei \cite{soma_13,soma_14}. An alternative method consists of using microscopic approaches based on the self-consistent mean-field 
theory and its extensions beyond mean-field. In nuclear physics, they are usually based on energy density functionals built from 
phenomenological parametrizations of the NN effective interaction, such as Skyrme \cite{vautherin_72,hellemans_13} or 
Gogny forces \cite{decharge_80,berger_91,chappert_08,goriely_09}. These approaches have successfully predicted a broad body of 
nuclear structure observables for nuclei ranging from medium to heavy masses. This wealth of developments can be extended to reaction 
calculations based on NN effective interaction. The so-called Nuclear Structure Method (NSM) for scattering 
\cite{vinhmau_70,vinhmau_76,bernard_79,bouyssy_81,osterfeld_81} relies on the self-consistent Hartree-Fock (HF) and 
Random-Phase Approximations (RPA) of the microscopic optical potential. The former is a mean-field potential and the latter is a 
polarization potential built from target-nucleus excitations. A strictly equivalent method, the continuum particle-vibration coupling using a Skyrme 
interaction, has been recently applied to neutron scattering from $^{16}$O \cite{mizuyama_12c}, but neglecting part of the residual 
interaction in the coupling vertices. Other approaches are in progress, where optical potential is approximated as the HF term plus 
the imaginary part of the uncorrelated particle-hole potential neglecting the collectivity of target excited states 
\cite{pilipenko_12,xu_14}. 

We report on optical potential calculations using NSM \cite{vinhmau_70}. Here the optical potential $V$ consists 
of two components,
\begin{equation}
 V = V^{HF} + \Delta V. \label{v}
\end{equation}
The HF potential, $V^{HF}$, is the major contribution to the real part of the optical potential. $V^{HF}$ is calculated in coordinate 
space to ensure the correct asymptotic behavior of single-particle states. It is non local and energy independent due to the nature 
of Gogny interaction, which is of finite range and energy independent, respectively. Rearrangement contributions stemming from the 
density-dependent term of the interaction are also taken into account. 

The second component of the potential in Eq.~\eqref{v} is
\begin{equation}
 \Delta V = V^{PP}+V^{RPA}-2V^{(2)}, \label{dv}
\end{equation}
which is complex, energy dependent and non local. Here $V^{PP}$ and $V^{RPA}$ are contributions from particle-particle and 
particle-hole correlations, respectively. The uncorrelated particle-hole contribution $V^{(2)}$ is accounted for once in $V^{PP}$,  
and twice in $V^{RPA}$. As a matter of fact, if two-body correlations are neglected in Eq.~\eqref{dv} for $V^{PP}$ and $V^{RPA}$, 
then $\Delta V$ reduces to $V^{(2)}$ as expected \cite{vinhmau_70}.

As mentioned in Ref.~\cite{bouyssy_81}, if one works with an NN effective interaction with a density-dependent term, such 
as Gogny or Skyrme forces, attention must be paid to correlations already accounted for in the interaction. Indeed, part of 
particle-particle correlations is already contained at the HF level as far as $\textrm{Re}[V^{PP}]$ is concerned. We thus use the same 
prescription as in Ref.~\cite{bernard_79}, omitting $\textrm{Re}[V^{PP}]$ while $\textrm{Im}[V^{PP}]$ is approximated by 
$\textrm{Im}\left[V^{(2)}\right]$. Then Eq.~\eqref{dv} becomes
\begin{equation}
 \Delta V = \textrm{Im}\left[V^{(2)}\right]+V^{RPA}-2V^{(2)}. \label{sig-approx}
\end{equation}
From now on, equations are presented omitting spin for simplicity. For nucleons with incident energy E, the RPA potential reads
\begin{eqnarray}
  V^{RPA}({\bf r,r'},E)&=& \sum_{N\neq 0} \sumint_{\lambda}
  \bigg[{{n_{\lambda}}\over{E-\varepsilon_{\lambda}+E_{N}-i\Gamma(E_{N})}} \nonumber\\
  &+&{{1-n_{\lambda}}\over{E-\varepsilon_{\lambda}-E_{N}+i\Gamma(E_{N})}}\bigg]\nonumber\\
  &\times& \Omega^{N}_{\lambda}(\textbf{r})\Omega^{N}_{\lambda}(\textbf{r'}), \label{vrpa}
\end{eqnarray}
where $n_{i}$ and $\varepsilon_{i}$ are occupation number and energy of the single-particle state $\phi_{i}$ in the HF 
field, respectively. Subscripts $p$, $h$ and $\lambda$ refer to the quantum number of particle, hole and the intermediate 
single-particle, respectively. $E_{N}$ and $\Gamma(E_{N})$ represent the energy and the width of the $N^{th}$ excited state 
of the target, respectively. Additionally 
\begin{eqnarray}
 \Omega^{N}_{\lambda}(\textbf{r}) = \sum_{(p,h)} \left[X^{N,(p,h)}F_{ph\lambda}(\textbf{r}) + Y^{N,(p,h)}F_{hp\lambda}(\textbf{r})\right],
\end{eqnarray}
where $X$ and $Y$ denote the usual RPA amplitudes and
\begin{equation}
 F_{ij\lambda}(\textbf{r}) = \int d^{3} \textbf{r}_{1} \phi^{*}_{i}(\textbf{r}_{1})v(\textbf{r},\textbf{r}_{1})\left[1-\textrm{\^P}\right]\phi_{\lambda}(\textbf{r})\phi_{j}(\textbf{r}_{1}),
\end{equation}
where $\textrm{\^P}$ is a particle-exchange operator and $v$ is the NN effective interaction. The particle-hole 
contribution reads
\begin{eqnarray}
  V^{(2)}({\bf r,r'},E) &=& \frac{1}{2}\sum_{ij} \sumint_{\lambda} \bigg[{{n_{i}(1-n_{j})n_{\lambda}}\over{E-\varepsilon_{\lambda}+E_{ij}-i\Gamma(E_{ij})}}\nonumber\\
  &+&{{n_{j}(1-n_{i})(1-n_{\lambda})}\over{E-\varepsilon_{\lambda}-E_{ij}+i\Gamma(E_{ij})}}\bigg]\nonumber\\
  &\times&  F_{ij\lambda}(\textbf{r})F^{*}_{ij\lambda}(\textbf{r}'), \label{vph}
\end{eqnarray}
with $E_{ij} = \varepsilon_{i}-\varepsilon_{j}$, the uncorrelated particle-hole energy. 

The description of target excitations has been obtained by solving the RPA/D1S equations in a harmonic oscillator basis, including 
fifteen major shells \cite{blaizot_77} and using the D1S Gogny interaction \cite{berger_91}. We account for RPA excited states with 
spin up to $J=8$, including both parities in order to achieve convergence of the cross section. The first 1$^{-}$ state given by RPA, 
containing the translational spurious mode, is removed. In order to avoid spurious modes in the uncorrelated particle-hole term, we 
approximate the 1$^{-}$ contribution in $V^{(2)}$ by half that of the 1$^{-}$ contribution in $V^{RPA}$. Coupling to excited states 
results in a number of poles in Eqs.~\eqref{vrpa} and \eqref{vph}. Moreover, fluctuations appear in the imaginary part of the 
potential whenever the energy $E-E_{N}$ matches a resonance energy of the intermediate single-particle state $\phi_{\lambda}$. The 
leading inelastic doorways are those containing single-particle resonances. Although the RPA/D1S method provides a good overall 
description of the spectroscopic properties of double-closed shell nuclei, couplings to two or more particle-hole states and to 
continuum states are neglected. The impact of these couplings is a strength redistribution that can be handled assigning a finite 
width $\Gamma(E_{N})$ to each RPA state. It has the effect of averaging in energy and smoothing the potential. The resulting potential 
can then be identified with an optical model \cite{feshbach_58}. A microscopic calculation of these widths is beyond 
the scope of the present study. We include them phenomenologically as an interpolation between reasonable values. $\Gamma(E_{N})$ 
takes the value of 2, 5, 15 and 50~MeV, for excitation energies of 20, 50, 100 and 200~MeV, respectively. The integro-differential 
Schr\"{o}dinger equation for elastic scattering is solved retaining the non-local structure of the potential \cite{raynal_98}. 
Moreover optical potential calculations yield shape elastic, reaction and total cross sections \cite{feshbach_58}. The compound-elastic cross 
section has to be added to shape elastic cross section and subtracted from reaction cross section before comparison with 
data \cite{feshbach_58}. In a first attempt, we use the compound-elastic contribution from Hauser-Feshbach calculations with TALYS 
code \cite{koning_08} using Koning-Delaroche global potential \cite{koning_03}. These considerations are particularly relevant for 
neutron scattering below 10~MeV.

\begin{figure}[t]
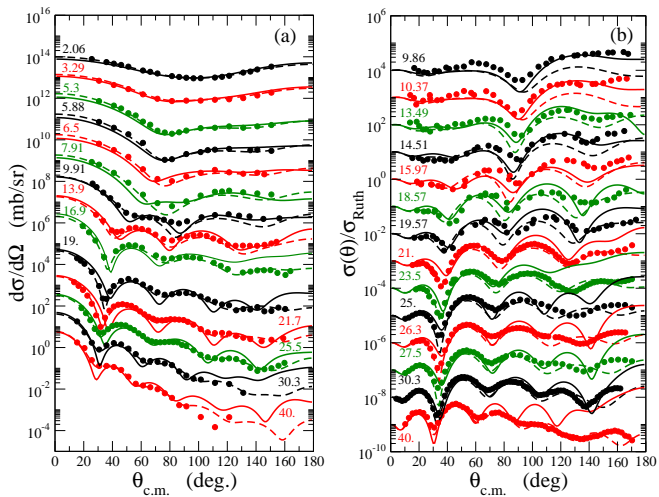

\begin{minipage}[c]{0.49\linewidth}
\begin{center}
\adjustbox{trim={0.\width} {0.\height} {0.\width} {0.\height},clip}
{\includegraphics[width=1.\textwidth,angle=-00,clip=false]{fig1.eps}}
\end{center}
\end{minipage}
\hfill
\begin{minipage}[c]{0.49\linewidth}
\begin{center}
\adjustbox{trim={0.\width} {0.\height} {0.\width} {0.\height},clip}
{\includegraphics[width=1.\textwidth,angle=-00,clip=right]{fig2.eps}}
\end{center}
\end{minipage}
\caption{Differential cross sections for neutron (a) and proton (b) scattering from $^{40}$Ca. Comparison between data 
(symbols), $V^{HF}+\Delta V$ results (solid curves) and Koning-Delaroche potential results (dashed curves).}
\label{sec-diff-rpaall}
\end{figure}

\begin{figure}[h]
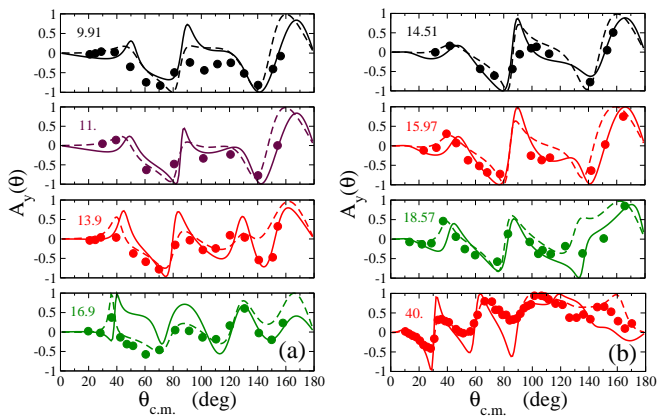

\begin{minipage}[c]{0.49\linewidth}
\begin{center}
\adjustbox{trim={0.\width} {0.\height} {0.\width} {0.\height},clip}
{\includegraphics[width=1.\textwidth,angle=-00,clip=false]{fig3.eps}}
\end{center}
\end{minipage}
\hfill
\begin{minipage}[c]{0.49\linewidth}
\begin{center}
\adjustbox{trim={0.\width} {0.\height} {0.\width} {0.\height},clip}
{\includegraphics[width=1.\textwidth,angle=-00,clip=false]{fig4.eps}}
\end{center}
\end{minipage}
\caption{Same as Fig.~\ref{sec-diff-rpaall} for analyzing powers.}
\label{pol-diff-rpaall}
\end{figure}

\begin{figure}[h]
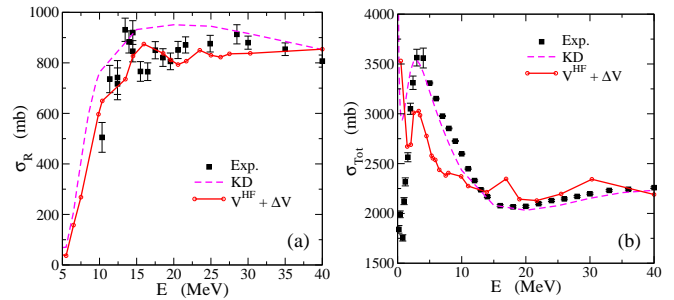

\begin{minipage}[c]{0.49\linewidth}
\begin{center}
\adjustbox{trim={0.\width} {0.\height} {0.\width} {0.\height},clip}
{\includegraphics[width=1.\textwidth,angle=-00,clip=false]{fig5.eps}}
\end{center}
\end{minipage}
\hfill
\begin{minipage}[c]{0.49\linewidth}
\begin{center}
\adjustbox{trim={0.\width} {0.\height} {0.\width} {0.\height},clip}
{\includegraphics[width=1.\textwidth,angle=-00,clip=false]{fig6.eps}}
\end{center}
\end{minipage}
\caption{Reaction cross section for proton (a) and total cross section for neutron (b) scattering from $^{40}$Ca. 
Comparison between data (symbols), $V^{HF}+\Delta V$ results (solid curve) and Koning-Delaroche potential (dashed curve).}
\label{cross-reac}
\end{figure}

In Fig.~\ref{sec-diff-rpaall}, we present results for the calculated differential cross sections based on NSM for both 
neutron and proton scattering from $^{40}$Ca. References to data are given in Ref.~\cite{koning_03}. Error bars are smaller than the size 
symbols. NSM results compare very well to experiment and those based on Koning-Delaroche potential up to about 30~MeV incident energy. 
Beyond 30~MeV, backward-angle cross sections are overestimated. Discrepancies at 16.9~MeV (23.5~MeV) for neutron (proton) 
scattering are related to resonances in the intermediate single-particle state when not completely averaged. A detailed treatment of the width 
might cure this issue. 
In Fig.~\ref{pol-diff-rpaall} we show calculated analyzing powers for neutron and proton scattering at several energies, in 
good agreement with measurements. Moreover, agreement with the data is comparable to that obtained from Koning-Delaroche 
potential. These results suggest that NSM potential retains the correct spin-orbit behavior.
In Fig.~\ref{cross-reac} we show reaction cross section for proton scattering (a) and total cross section for 
neutron scattering (b). Calculated reaction cross sections are in good agreement with experiments. For neutrons, 
however, we underestimate the total cross section below 10~MeV. Considering that the differential elastic cross section is well 
reproduced, this underestimate suggests that part of the absorption mechanism is not accounted for, as target-excited states beyond 
RPA or double-charge exchange process. 

\begin{figure}[h*]
\includegraphics[width=.3\textwidth,angle=-90,clip=false]{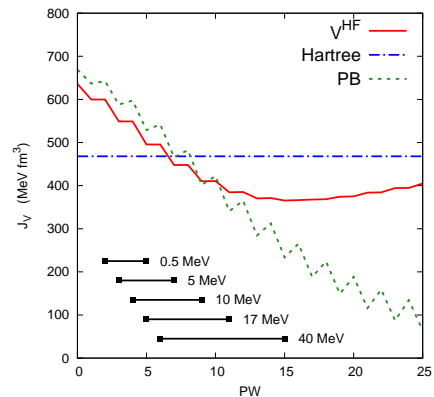}
\caption{Volume integral as a function of partial waves for neutron scattering from $^{40}$Ca: HF potential (solid curve), Hartree potential (dash-dotted 
curve) and Perey-Buck potential (dotted curve). Horizontal segments denote the partial-wave interval to sum up 80\% of the 
reaction cross section at selected incident energies.}
\label{fig:intvol_n}
\end{figure}
To understand the limited energy range of applicability of the NSM approach, we compare in Fig.~\ref{fig:intvol_n} the volume 
integral, $J_{V}$, of the central HF potential with the one obtained from the real part of the Perey-Buck non-local potential 
\cite{perey_62}. Black segments denote the strongest partial-wave contributions accounting for 80\% of the reaction 
cross section at the selected incident energies. Keep in mind that the HF potential is the leading contribution to the real part of $V$ 
in Eq. \eqref{v}. Its contribution to $J_{V}$ is similar to that from Perey-Buck up to about the twelfth partial 
wave ($\sim$17~MeV). Beyond this point HF saturates, following the trend of the Hartree potential which is local and thus partial-wave 
independent. This departure from Perey-Buck explains why increasing incident energy (partial wave) yields much too high $J_{V}$ for 
HF, with the subsequent overestimate of the differential cross section at backward angles. It would be interesting to investigate 
to what extent the effective interaction has incidence on this behavior at high partial wave.

\begin{figure}[h]
\begin{center}
\adjustbox{trim={0.\width} {0.\height} {0.\width} {0.\height},clip}
{\includegraphics[width=.3\textwidth,angle=-00,clip=false]{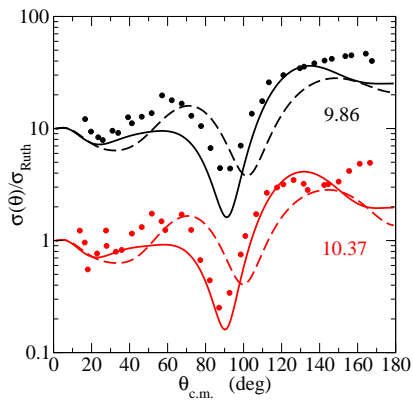}}
\end{center}
\vspace{-.5cm}
\caption{Differential cross sections $\sigma(\theta)/\sigma_{Ruth}$ for proton incident on $^{40}$Ca. Comparison between data 
(symbols), $V^{HF}~+~\Delta~V$ (solid curves) and $V^{HF}~+~\textrm{Im}~[V^{RPA}/2]$ results (dashed curves).}
\label{sec-imrpa2-rpaall}
\end{figure}

We now address the subtraction of the uncorrelated second-order term in Eq. \eqref{dv}. As pointed out in Ref.~\cite{barbieri_01}, 
this subtraction can lead to unphysical solutions with spurious poles and negative occupation numbers. The smooth and averaged 
potential obtained from Eq. \eqref{v} no longer suffers these pathologies. Indeed if one approximates $V^{(2)}~\approx~V^{RPA}/2$, then 
Eq. \eqref{dv} reduces to
\begin{equation}
 \Delta V \approx \textrm{Im}\left[V^{RPA}/2\right].\label{sig-approx2}
\end{equation}
This approximation has the drawback of neglecting the real part of $\Delta V$ as well as part of the collectivity of the excited states. 
However, it has the advantage of avoiding second-order double counting. 
\begin{figure}[h]
\begin{center}
\includegraphics[width=.4\textwidth,angle=-90,clip=false]{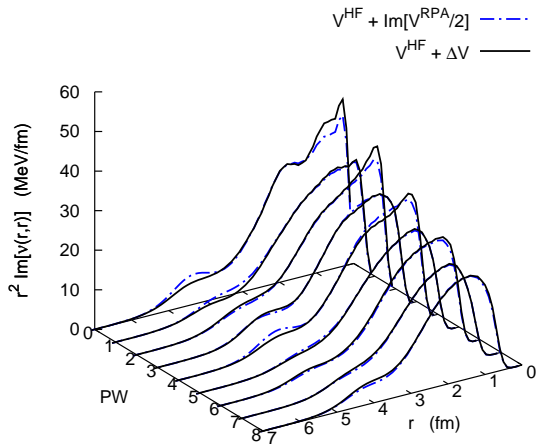}
\end{center}
\vspace{-1.cm}
\caption{Partial-wave contribution $\nu$ of  $\textrm{Im}~V$ for neutron scattering from $^{40}$Ca scattering at 9.91~MeV as a function of radius 
and partial waves: $V^{HF}+\Delta V$ potential (solid curve), $V^{HF}+\textrm{Im}[V^{RPA}/2]$ potential (dash-dotted curve).}
\label{fig:pw-40ca-n}
\end{figure}
As seen in Fig.~\ref{fig:pw-40ca-n}, both approximations in Eqs.~\eqref{sig-approx} and \eqref{sig-approx2} yield very similar shapes for each partial-wave 
contribution $\nu$ of the diagonal imaginary part of $\textrm{Im}[V]$ for neutron scattering from $^{40}$Ca at 9.91~MeV. This trend remains true for higher partial waves and 
incident energies, confirming the good behavior of $V^{HF}~+~\Delta~V$. In Fig.~\ref{sec-imrpa2-rpaall} we present the differential cross 
section for proton scattering from $^{40}$Ca based on these two approximations. The diffractive minima obtained with $V^{HF}~+~\Delta~V$ agree better 
with experiment than those obtained from $V^{HF}+ \textrm{Im}[V^{RPA}/2]$. This result emphasizes the important role played by the real part of $\Delta V$. 

The work presented here constitutes a promising step forward aimed at a model keeping at the same footing both reaction and structure 
aspects of the many-nucleon system. Within the optical model potential, NSM is able to account reasonably well for 
low energy scattering data. An important feature of the approach is the extraction of the imaginary part of the potential by means 
of intermediate excitations of the system. It has been based on Gogny effective interaction, although it can be applied to any interaction 
of similar nature. The study has been restricted to closed-shell targets but can be extended to deformed 
nuclei described with Quasi-particle RPA. Those results also open the way to new parametrizations of NN effective interactions including reaction 
phenomena. A comprehensive work on the formalism and applications shall be presented elsewhere.
\\
H. F. A. acknowledges partial funding from FONDECYT under Grant No 1120396.


\begin{thebibliography}{32}
\expandafter\ifx\csname natexlab\endcsname\relax\def\natexlab#1{#1}\fi
\expandafter\ifx\csname bibnamefont\endcsname\relax
  \def\bibnamefont#1{#1}\fi
\expandafter\ifx\csname bibfnamefont\endcsname\relax
  \def\bibfnamefont#1{#1}\fi
\expandafter\ifx\csname citenamefont\endcsname\relax
  \def\citenamefont#1{#1}\fi
\expandafter\ifx\csname url\endcsname\relax
  \def\url#1{\texttt{#1}}\fi
\expandafter\ifx\csname urlprefix\endcsname\relax\def\urlprefix{URL }\fi
\providecommand{\bibinfo}[2]{#2}
\providecommand{\eprint}[2][]{\url{#2}}

\bibitem[{\citenamefont{Hufner and Mahaux}(1972)}]{hufner_72}
\bibinfo{author}{\bibfnamefont{J.}~\bibnamefont{Hufner}} \bibnamefont{and}
  \bibinfo{author}{\bibfnamefont{C.}~\bibnamefont{Mahaux}},
  \bibinfo{journal}{Ann. Phys. (NY)} \textbf{\bibinfo{volume}{73}},
  \bibinfo{pages}{525 } (\bibinfo{year}{1972}).

\bibitem[{\citenamefont{Dupuis et~al.}(2006)\citenamefont{Dupuis, Karataglidis,
  Bauge, Delaroche, and Gogny}}]{dupuis_06}
\bibinfo{author}{\bibfnamefont{M.}~\bibnamefont{Dupuis}},
  \bibinfo{author}{\bibfnamefont{S.}~\bibnamefont{Karataglidis}},
  \bibinfo{author}{\bibfnamefont{E.}~\bibnamefont{Bauge}},
  \bibinfo{author}{\bibfnamefont{J.~P.} \bibnamefont{Delaroche}},
  \bibnamefont{and} \bibinfo{author}{\bibfnamefont{D.}~\bibnamefont{Gogny}},
  \bibinfo{journal}{Phys. Rev. C} \textbf{\bibinfo{volume}{73}},
  \bibinfo{pages}{014605} (\bibinfo{year}{2006}).

\bibitem[{\citenamefont{Arellano and von Geramb}(2002)}]{arellano_02}
\bibinfo{author}{\bibfnamefont{H.~F.} \bibnamefont{Arellano}} \bibnamefont{and}
  \bibinfo{author}{\bibfnamefont{H.~V.} \bibnamefont{von Geramb}},
  \bibinfo{journal}{Phys. Rev. C} \textbf{\bibinfo{volume}{66}},
  \bibinfo{pages}{024602} (\bibinfo{year}{2002}).

\bibitem[{\citenamefont{Quaglioni and Navr\'atil}(2008)}]{quaglioni_08}
\bibinfo{author}{\bibfnamefont{S.}~\bibnamefont{Quaglioni}} \bibnamefont{and}
  \bibinfo{author}{\bibfnamefont{P.}~\bibnamefont{Navr\'atil}},
  \bibinfo{journal}{Phys. Rev. Lett.} \textbf{\bibinfo{volume}{101}},
  \bibinfo{pages}{092501} (\bibinfo{year}{2008}).

\bibitem[{\citenamefont{Hupin et~al.}(2013)\citenamefont{Hupin, Langhammer,
  Navr\'atil, Quaglioni, Calci, and Roth}}]{hupin_13}
\bibinfo{author}{\bibfnamefont{G.}~\bibnamefont{Hupin}},
  \bibinfo{author}{\bibfnamefont{J.}~\bibnamefont{Langhammer}},
  \bibinfo{author}{\bibfnamefont{P.}~\bibnamefont{Navr\'atil}},
  \bibinfo{author}{\bibfnamefont{S.}~\bibnamefont{Quaglioni}},
  \bibinfo{author}{\bibfnamefont{A.}~\bibnamefont{Calci}}, \bibnamefont{and}
  \bibinfo{author}{\bibfnamefont{R.}~\bibnamefont{Roth}},
  \bibinfo{journal}{Phys. Rev. C} \textbf{\bibinfo{volume}{88}},
  \bibinfo{pages}{054622} (\bibinfo{year}{2013}).

\bibitem[{\citenamefont{Nollett et~al.}(2007)\citenamefont{Nollett, Pieper,
  Wiringa, Carlson, and Hale}}]{nollett_07}
\bibinfo{author}{\bibfnamefont{K.~M.} \bibnamefont{Nollett}},
  \bibinfo{author}{\bibfnamefont{S.~C.} \bibnamefont{Pieper}},
  \bibinfo{author}{\bibfnamefont{R.~B.} \bibnamefont{Wiringa}},
  \bibinfo{author}{\bibfnamefont{J.}~\bibnamefont{Carlson}}, \bibnamefont{and}
  \bibinfo{author}{\bibfnamefont{G.~M.} \bibnamefont{Hale}},
  \bibinfo{journal}{Phys. Rev. Lett.} \textbf{\bibinfo{volume}{99}},
  \bibinfo{pages}{022502} (\bibinfo{year}{2007}).

\bibitem[{\citenamefont{Dussan et~al.}(2011)\citenamefont{Dussan, Waldecker,
  Dickhoff, M\"uther, and Polls}}]{dussan_11}
\bibinfo{author}{\bibfnamefont{H.}~\bibnamefont{Dussan}},
  \bibinfo{author}{\bibfnamefont{S.~J.} \bibnamefont{Waldecker}},
  \bibinfo{author}{\bibfnamefont{W.~H.} \bibnamefont{Dickhoff}},
  \bibinfo{author}{\bibfnamefont{H.}~\bibnamefont{M\"uther}}, \bibnamefont{and}
  \bibinfo{author}{\bibfnamefont{A.}~\bibnamefont{Polls}},
  \bibinfo{journal}{Phys. Rev. C} \textbf{\bibinfo{volume}{84}},
  \bibinfo{pages}{044319} (\bibinfo{year}{2011}).

\bibitem[{\citenamefont{Barbieri and Jennings}(2005)}]{barbieri_05}
\bibinfo{author}{\bibfnamefont{C.}~\bibnamefont{Barbieri}} \bibnamefont{and}
  \bibinfo{author}{\bibfnamefont{B.~K.} \bibnamefont{Jennings}},
  \bibinfo{journal}{Phys. Rev. C} \textbf{\bibinfo{volume}{72}},
  \bibinfo{pages}{014613} (\bibinfo{year}{2005}).

\bibitem[{\citenamefont{Hagen and Michel}(2012)}]{hagen_12}
\bibinfo{author}{\bibfnamefont{G.}~\bibnamefont{Hagen}} \bibnamefont{and}
  \bibinfo{author}{\bibfnamefont{N.}~\bibnamefont{Michel}},
  \bibinfo{journal}{Phys. Rev. C} \textbf{\bibinfo{volume}{86}},
  \bibinfo{pages}{021602} (\bibinfo{year}{2012}).

\bibitem[{\citenamefont{Som\`a et~al.}(2013)\citenamefont{Som\`a, Barbieri, and
  Duguet}}]{soma_13}
\bibinfo{author}{\bibfnamefont{V.}~\bibnamefont{Som\`a}},
  \bibinfo{author}{\bibfnamefont{C.}~\bibnamefont{Barbieri}}, \bibnamefont{and}
  \bibinfo{author}{\bibfnamefont{T.}~\bibnamefont{Duguet}},
  \bibinfo{journal}{Phys. Rev. C} \textbf{\bibinfo{volume}{87}},
  \bibinfo{pages}{011303} (\bibinfo{year}{2013}).

\bibitem[{\citenamefont{Som\`a et~al.}(2014)\citenamefont{Som\`a, Barbieri, and
  Duguet}}]{soma_14}
\bibinfo{author}{\bibfnamefont{V.}~\bibnamefont{Som\`a}},
  \bibinfo{author}{\bibfnamefont{C.}~\bibnamefont{Barbieri}}, \bibnamefont{and}
  \bibinfo{author}{\bibfnamefont{T.}~\bibnamefont{Duguet}},
  \bibinfo{journal}{Phys. Rev. C} \textbf{\bibinfo{volume}{89}},
  \bibinfo{pages}{024323} (\bibinfo{year}{2014}).

\bibitem[{\citenamefont{Vautherin and Brink}(1972)}]{vautherin_72}
\bibinfo{author}{\bibfnamefont{D.}~\bibnamefont{Vautherin}} \bibnamefont{and}
  \bibinfo{author}{\bibfnamefont{D.~M.} \bibnamefont{Brink}},
  \bibinfo{journal}{Phys. Rev. C} \textbf{\bibinfo{volume}{5}},
  \bibinfo{pages}{626} (\bibinfo{year}{1972}).

\bibitem[{\citenamefont{Hellemans et~al.}(2013)\citenamefont{Hellemans,
  Pastore, Duguet, Bennaceur, Davesne, Meyer, Bender, and
  Heenen}}]{hellemans_13}
\bibinfo{author}{\bibfnamefont{V.}~\bibnamefont{Hellemans}},
  \bibinfo{author}{\bibfnamefont{A.}~\bibnamefont{Pastore}},
  \bibinfo{author}{\bibfnamefont{T.}~\bibnamefont{Duguet}},
  \bibinfo{author}{\bibfnamefont{K.}~\bibnamefont{Bennaceur}},
  \bibinfo{author}{\bibfnamefont{D.}~\bibnamefont{Davesne}},
  \bibinfo{author}{\bibfnamefont{J.}~\bibnamefont{Meyer}},
  \bibinfo{author}{\bibfnamefont{M.}~\bibnamefont{Bender}}, \bibnamefont{and}
  \bibinfo{author}{\bibfnamefont{P.-H.} \bibnamefont{Heenen}},
  \bibinfo{journal}{Phys. Rev. C} \textbf{\bibinfo{volume}{88}},
  \bibinfo{pages}{064323} (\bibinfo{year}{2013}).

\bibitem[{\citenamefont{Decharg\'e and Gogny}(1980)}]{decharge_80}
\bibinfo{author}{\bibfnamefont{J.}~\bibnamefont{Decharg\'e}} \bibnamefont{and}
  \bibinfo{author}{\bibfnamefont{D.}~\bibnamefont{Gogny}},
  \bibinfo{journal}{Phys. Rev. C} \textbf{\bibinfo{volume}{21}},
  \bibinfo{pages}{1568} (\bibinfo{year}{1980}).

\bibitem[{\citenamefont{Berger et~al.}(1991)\citenamefont{Berger, Girod, and
  Gogny}}]{berger_91}
\bibinfo{author}{\bibfnamefont{J.~F.} \bibnamefont{Berger}},
  \bibinfo{author}{\bibfnamefont{M.}~\bibnamefont{Girod}}, \bibnamefont{and}
  \bibinfo{author}{\bibfnamefont{D.}~\bibnamefont{Gogny}},
  \bibinfo{journal}{Comput. Phys. Commun.} \textbf{\bibinfo{volume}{63}},
  \bibinfo{pages}{365} (\bibinfo{year}{1991}).

\bibitem[{\citenamefont{Chappert et~al.}(2008)\citenamefont{Chappert, Girod,
  and Hilaire}}]{chappert_08}
\bibinfo{author}{\bibfnamefont{F.}~\bibnamefont{Chappert}},
  \bibinfo{author}{\bibfnamefont{M.}~\bibnamefont{Girod}}, \bibnamefont{and}
  \bibinfo{author}{\bibfnamefont{S.}~\bibnamefont{Hilaire}},
  \bibinfo{journal}{Phys. Lett. B} \textbf{\bibinfo{volume}{668}},
  \bibinfo{pages}{420 } (\bibinfo{year}{2008}).

\bibitem[{\citenamefont{Goriely et~al.}(2009)\citenamefont{Goriely, Hilaire,
  Girod, and P\'eru}}]{goriely_09}
\bibinfo{author}{\bibfnamefont{S.}~\bibnamefont{Goriely}},
  \bibinfo{author}{\bibfnamefont{S.}~\bibnamefont{Hilaire}},
  \bibinfo{author}{\bibfnamefont{M.}~\bibnamefont{Girod}}, \bibnamefont{and}
  \bibinfo{author}{\bibfnamefont{S.}~\bibnamefont{P\'eru}},
  \bibinfo{journal}{Phys. Rev. Lett.} \textbf{\bibinfo{volume}{102}},
  \bibinfo{pages}{242501} (\bibinfo{year}{2009}).

\bibitem[{\citenamefont{Vinh~Mau}(1970)}]{vinhmau_70}
\bibinfo{author}{\bibfnamefont{N.}~\bibnamefont{Vinh~Mau}},
  \bibinfo{journal}{Theory of nuclear structure (IAEA, Vienna)} p.
  \bibinfo{pages}{931} (\bibinfo{year}{1970}).

\bibitem[{\citenamefont{Vinh~Mau and Bouyssy}(1976)}]{vinhmau_76}
\bibinfo{author}{\bibfnamefont{N.}~\bibnamefont{Vinh~Mau}} \bibnamefont{and}
  \bibinfo{author}{\bibfnamefont{A.}~\bibnamefont{Bouyssy}},
  \bibinfo{journal}{Nucl. Phys.} \textbf{\bibinfo{volume}{A257}},
  \bibinfo{pages}{189 } (\bibinfo{year}{1976}), ISSN \bibinfo{issn}{0375-9474}.

\bibitem[{\citenamefont{Bernard and Van~Giai}(1979)}]{bernard_79}
\bibinfo{author}{\bibfnamefont{V.}~\bibnamefont{Bernard}} \bibnamefont{and}
  \bibinfo{author}{\bibfnamefont{N.}~\bibnamefont{Van~Giai}},
  \bibinfo{journal}{Nucl. Phys.} \textbf{\bibinfo{volume}{A327}},
  \bibinfo{pages}{397 } (\bibinfo{year}{1979}).

\bibitem[{\citenamefont{Bouyssy et~al.}(1981)\citenamefont{Bouyssy, Ng{\^o},
  and Vinh~Mau}}]{bouyssy_81}
\bibinfo{author}{\bibfnamefont{A.}~\bibnamefont{Bouyssy}},
  \bibinfo{author}{\bibfnamefont{H.}~\bibnamefont{Ng{\^o}}}, \bibnamefont{and}
  \bibinfo{author}{\bibfnamefont{N.}~\bibnamefont{Vinh~Mau}},
  \bibinfo{journal}{Nucl. Phys.} \textbf{\bibinfo{volume}{A371}},
  \bibinfo{pages}{173 } (\bibinfo{year}{1981}).

\bibitem[{\citenamefont{Osterfeld et~al.}(1981)\citenamefont{Osterfeld,
  Wambach, and Madsen}}]{osterfeld_81}
\bibinfo{author}{\bibfnamefont{F.}~\bibnamefont{Osterfeld}},
  \bibinfo{author}{\bibfnamefont{J.}~\bibnamefont{Wambach}}, \bibnamefont{and}
  \bibinfo{author}{\bibfnamefont{V.~A.} \bibnamefont{Madsen}},
  \bibinfo{journal}{Phys. Rev. C} \textbf{\bibinfo{volume}{23}},
  \bibinfo{pages}{179} (\bibinfo{year}{1981}).

\bibitem[{\citenamefont{Mizuyama and Ogata}(2012)}]{mizuyama_12c}
\bibinfo{author}{\bibfnamefont{K.}~\bibnamefont{Mizuyama}} \bibnamefont{and}
  \bibinfo{author}{\bibfnamefont{K.}~\bibnamefont{Ogata}},
  \bibinfo{journal}{Phys. Rev. C} \textbf{\bibinfo{volume}{86}},
  \bibinfo{pages}{041603} (\bibinfo{year}{2012}).

\bibitem[{\citenamefont{Pilipenko and Kuprikov}(2012)}]{pilipenko_12}
\bibinfo{author}{\bibfnamefont{V.~V.} \bibnamefont{Pilipenko}}
  \bibnamefont{and} \bibinfo{author}{\bibfnamefont{V.~I.}
  \bibnamefont{Kuprikov}}, \bibinfo{journal}{Phys. Rev. C}
  \textbf{\bibinfo{volume}{86}}, \bibinfo{pages}{064613}
  (\bibinfo{year}{2012}).

\bibitem[{\citenamefont{Xu et~al.}(2014)\citenamefont{Xu, Guo, Han, and
  Shen}}]{xu_14}
\bibinfo{author}{\bibfnamefont{Y.}~\bibnamefont{Xu}},
  \bibinfo{author}{\bibfnamefont{H.}~\bibnamefont{Guo}},
  \bibinfo{author}{\bibfnamefont{Y.}~\bibnamefont{Han}}, \bibnamefont{and}
  \bibinfo{author}{\bibfnamefont{Q.}~\bibnamefont{Shen}}, \bibinfo{journal}{J.
  Phys. G: Nucl. Part. Phys.} \textbf{\bibinfo{volume}{41}},
  \bibinfo{pages}{015101} (\bibinfo{year}{2014}).

\bibitem[{\citenamefont{Blaizot and Gogny}(1977)}]{blaizot_77}
\bibinfo{author}{\bibfnamefont{J.}~\bibnamefont{Blaizot}} \bibnamefont{and}
  \bibinfo{author}{\bibfnamefont{D.}~\bibnamefont{Gogny}},
  \bibinfo{journal}{Nucl. Phys.} \textbf{\bibinfo{volume}{A284}},
  \bibinfo{pages}{429 } (\bibinfo{year}{1977}).

\bibitem[{\citenamefont{Feshbach}(1958)}]{feshbach_58}
\bibinfo{author}{\bibfnamefont{H.}~\bibnamefont{Feshbach}},
  \bibinfo{journal}{Annals of Physics} \textbf{\bibinfo{volume}{5}},
  \bibinfo{pages}{357 } (\bibinfo{year}{1958}).

\bibitem[{\citenamefont{Raynal}()}]{raynal_98}
\bibinfo{author}{\bibfnamefont{J.}~\bibnamefont{Raynal}},
  \bibinfo{note}{computer code DWBA98, 1998, (NEA 1209/05)}.

\bibitem[{\citenamefont{Koning et~al.}(2008)\citenamefont{Koning, Hilaire, and
  Duijvestijn}}]{koning_08}
\bibinfo{author}{\bibfnamefont{A.~J.} \bibnamefont{Koning}},
  \bibinfo{author}{\bibfnamefont{S.}~\bibnamefont{Hilaire}}, \bibnamefont{and}
  \bibinfo{author}{\bibfnamefont{M.}~\bibnamefont{Duijvestijn}}, in
  \emph{\bibinfo{booktitle}{Proceeding of the International Conference on
  Nuclear\\ Data for Science and Technology-ND2007}} (\bibinfo{publisher}{EDP
  Sciences}, \bibinfo{address}{Paris, France}, \bibinfo{year}{2008}), pp.
  \bibinfo{pages}{211--214}.

\bibitem[{\citenamefont{Koning and Delaroche}(2003)}]{koning_03}
\bibinfo{author}{\bibfnamefont{A.~J.} \bibnamefont{Koning}} \bibnamefont{and}
  \bibinfo{author}{\bibfnamefont{J.~P.} \bibnamefont{Delaroche}},
  \bibinfo{journal}{Nucl. Phys.} \textbf{\bibinfo{volume}{A713}},
  \bibinfo{pages}{231 } (\bibinfo{year}{2003}).

\bibitem[{\citenamefont{Perey and Buck}(1962)}]{perey_62}
\bibinfo{author}{\bibfnamefont{F.}~\bibnamefont{Perey}} \bibnamefont{and}
  \bibinfo{author}{\bibfnamefont{B.}~\bibnamefont{Buck}},
  \bibinfo{journal}{Nucl. Phys.} \textbf{\bibinfo{volume}{32}},
  \bibinfo{pages}{353 } (\bibinfo{year}{1962}).

\bibitem[{\citenamefont{Barbieri and Dickhoff}(2001)}]{barbieri_01}
\bibinfo{author}{\bibfnamefont{C.}~\bibnamefont{Barbieri}} \bibnamefont{and}
  \bibinfo{author}{\bibfnamefont{W.~H.} \bibnamefont{Dickhoff}},
  \bibinfo{journal}{Phys. Rev. C} \textbf{\bibinfo{volume}{63}},
  \bibinfo{pages}{034313} (\bibinfo{year}{2001}).

\end{thebibliography}

\end{document}